\documentclass[preprint]{aastex}

\newcommand{\kms} {km\,s$^{-1}$}
\newcommand{\cm}{cm$^{-1}$}
\newcommand{\ecar}{$ \eta\ Car$}
\newcommand{\conf}[3]{#1#2$\!^\mathrm{#3}$}
\newcommand{\term}[3]{#1\,$^\mathrm{#2}$#3}

\begin{document}

\shorttitle{Eta Carinae Ejecta} \shortauthors{Gull et al.}

\title{The Absorption Spectrum of High-Density Stellar Ejecta in
the Line-of-Sight to Eta Carinae}
\author{ T.~R.~Gull\altaffilmark{1}, G.~Vieira\altaffilmark{1,2},
F.~Bruhweiler\altaffilmark{1,3}, K.E.~Nielsen\altaffilmark{1,3},
E.~Verner\altaffilmark{1,3},A.~Danks\altaffilmark{4}}

\altaffiltext{1}{Laboratory for Astronomy and Solar Physics, Code
681, Goddard Space Flight Center, Greenbelt, MD 20771}
\altaffiltext{2}{Science Systems and Applications, Inc, Lanham, MD
20706} \altaffiltext{3} {Dept. of Physics/IACS, Catholic University of America,
Washington, DC 20064}\altaffiltext{4}{SGT - Inc, Greenbelt, MD
20770}

\email{Theodore.R.Gull@nasa.gov, GVieira@stis.gsfc.nasa.gov,
FredB@iacs.gsfc.nasa.gov, Nielsen@stis.gsfc.nasa.gov,
KVerner@stis.gsfc.nasa.gov, Danks@ngc.com}

\begin{abstract}
Using the high dispersion NUV mode of the Space Telescope Imaging
Spectrograph (STIS) aboard the Hubble Space Telescope (HST) to
observe Eta Carinae, we have resolved and identified over 500
sharp, circumstellar absorption lines of iron-group singly-ionized
and neutral elements with $\approx$20 velocity components ranging
from $-$146 \kms{} to $-$585 \kms{}. These lines are from
transitions originating from ground and metastable levels as high
as 40,000 cm$^{-1}$ above ground. The absorbing material is located either in
dense inhomogeneities in the stellar wind, the warm circumstellar
gas immediately in the vicinity of Eta Carinae, or within the
cooler foreground lobe of the Homunculus. We have used classical
curve-of-growth analysis to derive atomic level populations for \ion{Fe} {2} at $-$146 \kms{} and for \ion{Ti}{2} at
$-$513 \kms{}. These populations, plus photoionization and
statistical equilibrium modeling, provide electron temperatures, T$_\mathrm{e}$, densities, n$_\mathrm{H}$, and constraints on distances from the stellar source,
r. For the $-$146 km~s$^{-1}$ component, we derive T$_\mathrm{e}$ = 6400 K,
n$_\mathrm{H} \ge$ 10$^7 -$ 10$^8$ cm$^{-3}$, and d $\approx$ 1300 AU. For the
$-$513 km~s$^{-1}$ component, we find a much cooler temperature,
T$_\mathrm{e}$= 760 K,  with n$_\mathrm{H} \ge$  10$^7$ cm$^{-3}$, we estimate d$\approx$
10,000 AU. The large distances for these two components place the
absorptions in the vicinity of identifiable ejecta from historical
events, not near  or in the dense wind
of \ecar. Further analysis, in parallel with obtaining improved
experimental and theoretical atomic data, is underway to determine
what physical mechanisms and elemental abundances can explain the
large number of strong circumstellar absorption features in the
spectrum of \ecar.

\end{abstract}

\section{Introduction}
Eta Carinae (\ecar) has been an enigma since the Great Outburst in
the 1840's (Davidson \& Humphreys 1999). The 100+ M$_\sun$ star,
located at a distance $\sim$2300 pc (Meaburn 1999), with
$\sim$10$^\mathrm{6.7}$L$_\sun$ and a very large mass loss rate of
$\sim$10$^\mathrm{-3}$M$_\sun $y$^{-1}$ is surrounded by a dusty,
expanding double-lobed nebular complex known as the Homunculus.
Smith et al. (2003) using infrared images estimated that the
Homunculus may contain 12 M$_\sun$  of material, most having been
ejected since the Great Outburst. Because of its recent activity,
proximity, and brightness, \ecar{} offers a unique opportunity to
study mass loss processes of a massive star in considerable
detail. Since massive Population III stars are thought to have
been abundant in the Early Universe (Heger \& Woosley 2003), the
nebulosities of \ecar{} may provide fundamental clues to how
extensive mass loss in the last stages of evolution in extremely
massive stars, both now and in the Early Universe, enriches the
interstellar medium in processed elements.

The observations described herein clearly indicate a rich
circumstellar absorption spectrum for the ejecta of \ecar. These
narrow, interstellar-like features are seen in roughly 20
components with heliocentric velocities from $-$146 to $-$584
km~s$^{-1}$. The spectral lines of this absorption complex cannot
be interstellar and arise in gas {\it at much higher densities} (n
$\approx$ 10$^{7}$ cm$^{-3}$) than those of the multiple
interstellar components seen from $+$100 to $-$50 \kms{} toward
other stars in the Carina association (Walborn et al. 2002). The
interstellar lines, typifying densities n $\leq$ 10$^{4}$
cm$^{-3}$, are  formed in the foreground, much lower density H~II
region gas and interstellar clouds along the
line-of-sight to \ecar.

\section{Observations and Data Reduction}
Observations were obtained with the HST/STIS  using the E230H
grating and the $0\,\farcs2 \times 0\,\farcs09$ aperture,
providing a spectral resolving power of R$\sim$114,000 in the near
UV. The HST high spatial resolution, about $0\,\farcs03$, provides
critical separation of the stellar emission from  dust-scattered
starlight and nebular emission within the Homunculus.  The chaotic
nature of the expanding ejecta within the Homunculus, the extended
wind of the central source and the many  emission nebulosities,
lead to a velocity-smeared spectrum for  spatial scales exceeding
$0\,\farcs 1$. These two-dimensional  STIS spectral images were
acquired under  STIS GTO program 9242 on October 1, 2001 (E230H,
2385 to 2940 \AA) and GO program 9083 on January 20, 2002 (E230H,
2885 to 3160 \AA). Data reduction was performed with the CALSTIS
software (Lindler 1999) developed at NASA/GSFC.

\section{Analysis}
\begin{figure}[ht]
\scalebox{0.65}{\rotatebox{90}{\plotone{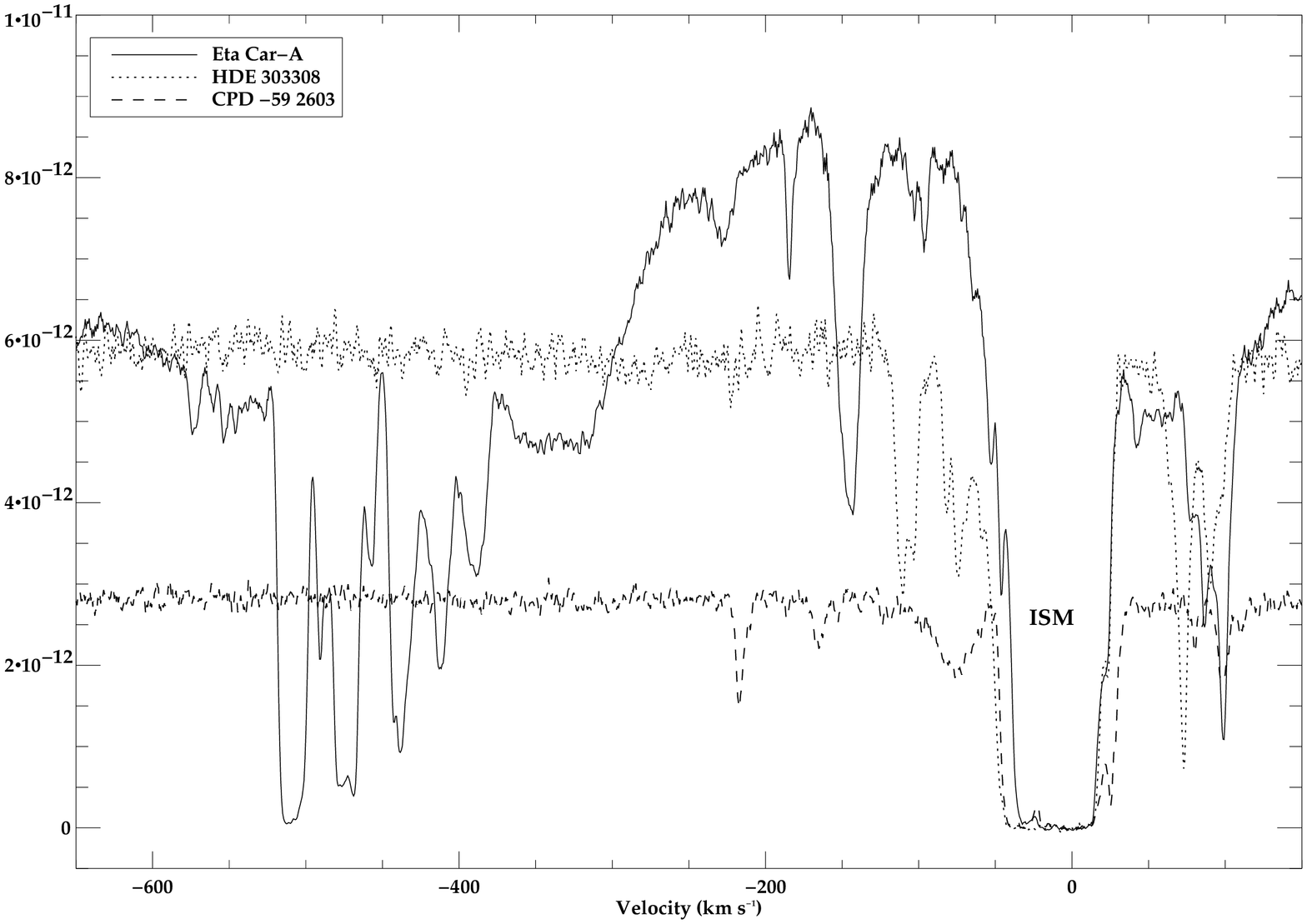}}} \caption{Spectral
region of \ion{Mg}{1} $\lambda$2853 for \ecar{}, HDE\,303308 and
CPD\,$-$59\degr\,2603, all members of the Carinae association.
Strong intervening IS absorption velocities are $+$100 to $-$50
\kms{} (Walborn et al. 2002). Unique to the \ecar\ spectrum are
the prominent absorption line at $-$146 \kms{} and the multiple,
narrow absorption lines between $-$385 and $-$585 \kms{}. The
\ecar{} spectrum includes a slowly varying P-Cygni profile in
addition to a series of narrower absorption components between
$-$146 and $-$585 \kms{}. \label{fig1}}
\end{figure}

\begin{figure}[ht]
\scalebox{0.7}  {\includegraphics{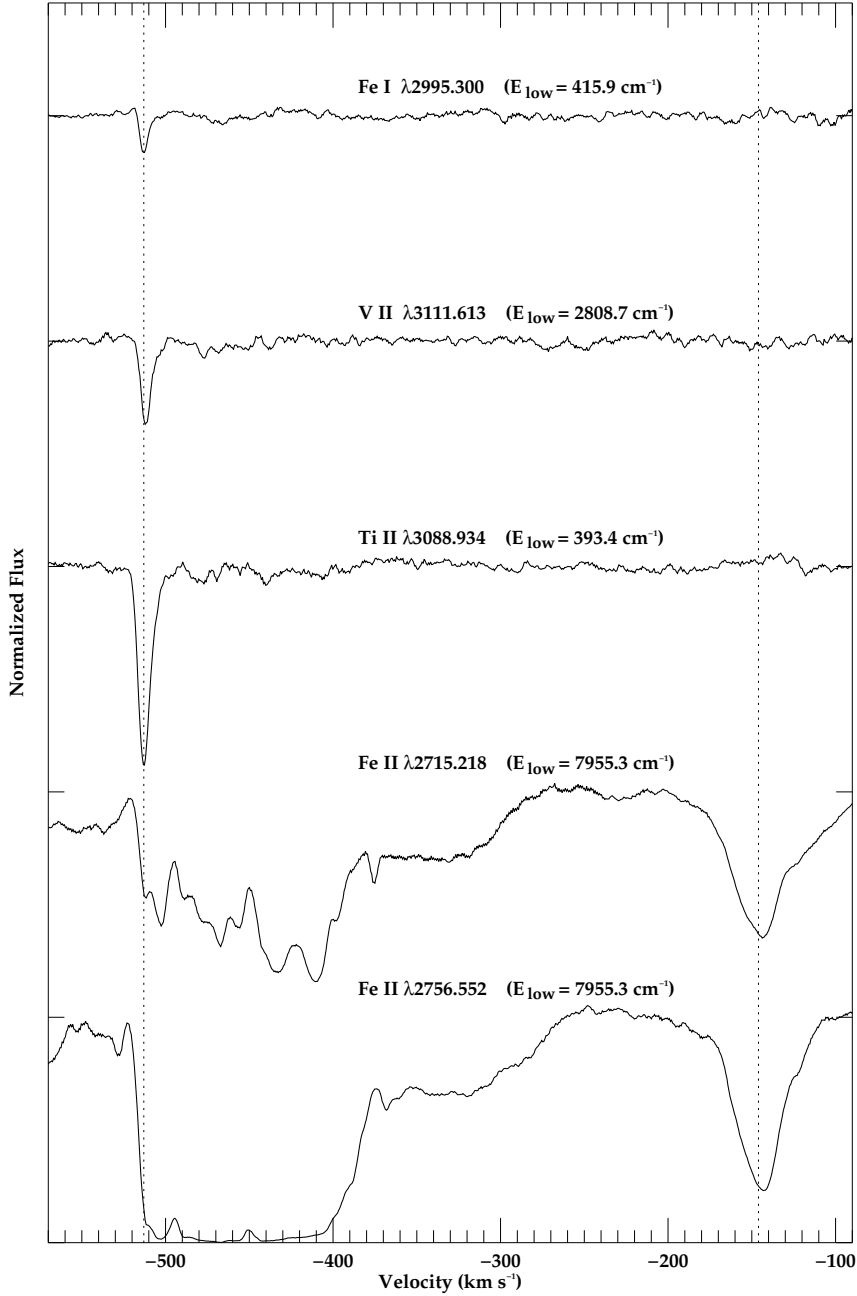}} \caption{Comparison of
circumstellar absorption line profiles for selected observed
transitions. These normalized spectral profiles are for
transitions from a range of lower levels for several neutral and
singly-ionized elements. The $-$513 \kms{} component is the
narrowest, yet absorption lines originate from levels up to 7400
\cm above ground. Note that many \ion{Fe}{1}, \ion{V}{2} and
\ion{Ti}{2} lines are unique to the $-$513 \kms{} component. The
$-$146 \kms{} component is from singly-ionized, iron-peak and
other elements, includes transitions from even higher energy
levels and is characteristically broader than the lines of the
$-$513 \kms{} component. The complicated velocity structure
between $-$380 to $-$513 \kms{} is noticeable for the two
\ion{Fe}{2} lines. The two \ion{Fe}{2} lines also have a very
broad wind component, starting from $-$250 \kms{} and extending
into the portion of the spectrum of the narrow line components.
Spatial resolution of these structures demonstrate that the wind
feature is very distinct from the narrow velocity components.
\label{fig2}}
\end{figure}

The STIS spectrum  of \ecar{} displays  sharp circumstellar and interstellar (IS)
lines formed along the line-of-sight (Figure~\ref{fig1}). Walborn
et al. (2002) identified multiple IS components in the direction
of four association stars (including HDE\,303308 and
CPD\,$-$59$\degr$2603), ranging in velocity from $+$127 to $-$388
\kms{}. The strongest IS lines, all arising from levels within the
ground configuration of the observed species with velocities
ranging from $+$100 to $-$50 \kms{}, are also present in the
\ecar{} spectrum. Unique to the \ecar{} spectrum are absorption
lines at $-$146 \kms{} and the complex circumstellar absorptions
between $-$385 to $-$585 \kms{} coming from a wide range of
metastable levels, up to 40,000 cm$^{-1}$, in neutral and singly
ionized species. Because of the high level of excitation, these
absorptions imply densities far above what is expected in the
normal ISM. In Figure~\ref{fig2}, absorption line profiles are
plotted from $-$100 to $-$570 \kms{} for selected lines of various
species. The slowly varying stellar flux is normalized to
facilitate the comparison. Two velocity components are readily
apparent: one at $-$146 \kms{} and another at $-$513 \kms{}. In
what follows, we will devote our analysis to Fe II and \ion{Ti}{2},
respectively, in these two velocity components since they are
fairly isolated with minimal contamination by absorption from
adjacent components.

\indent The two absorption components at $-$146 and $-$513 \kms\  are easily distinguishable
by their very different characteristic widths and by the range in
originating energy levels. The relatively broad component at
$-$146 \kms{}, observed in lines of \ion{Mg}{2}, \ion{Cr}{2},
\ion{Fe}{2}, \ion{Mn}{2}, \ion{Ni}{2} and \ion{Co}{2}, originates
from a wide range of energy levels. In Figure 2, the $-$146 \kms{}
component is present in \ion{Fe}{2} $\lambda\lambda$2715, 2756,
both lines arising from a level 7955 cm$^{-1}$ above ground. It is
significantly broader than the multiple, blended components
between $-$385 and $-$585 \kms{}. This component, with identified
lines originating from energy levels up to 40,000 cm$^{-1}$(nearly
5 eV above ground!), is more excited than the higher velocity
components.

\indent The velocity component at $-$513 \kms{} is observed in
lines of both neutral and singly ionized species as  in
\ion{Fe}{1} $\lambda$2995, \ion{V}{2} $\lambda$3111 and
\ion{Ti}{2} $\lambda$3088 (Figure 2). While we have identified
weak \ion{Fe}{1} lines originating from levels as high as 7400
cm$^{-1}$, nearly all lines originate from much lower levels. The
\ion{V}{2} absorption lines originate from levels ranging up to
3200 \cm ($\sim$0.4 eV above ground). So far, only upper limits
have been found for  \ion{V}{2} in the interstellar medium
(Cardelli 1994; Snow, Weiler \& Oegerle 1984; Welty et al. 1992),
where these results imply vanadium is at least thirty times under
abundant compared to solar values. In the $-$513 \kms{} absorption
component in the spectrum of \ecar, we have unambiguously detected
over twenty strong absorption lines of \ion{V}{2}. The difference
in ionization and the relative level populations between the
$-$146 and the $-$513 \kms{}  velocity components indicate these
components originate from regions of very different ionization and
temperature, and at very different distances from \ecar{}.

\indent Extreme care was taken in extracting the information
necessary to derive ionic column densities in the $-$146 \kms{}
and $-$513 \kms{} components. Measuring reliable equivalent widths
was complicated by a) absorption from other components at
velocities near the component of interest, and b) an often highly
undulating local continuum due to the complex spectrum produced by
the underlying photospheric and wind features of the central star.
Where the underlying local continuum is flat, the equivalent
widths and velocity centroids are reliable. But, when the
local continuum varies sharply with wavelength, the line profile
information can be less well defined. Many lines that are
obviously present are hopelessly blended or have ill-defined
continua; we therefor rejected measures of these lines.
Fortunately, there are enough strong, well-isolated lines with
high signal-to-noise to derive useful physical parameters.

\indent The derived equivalent widths have been analyzed using
standard curve-of-growth techniques to determine  velocity
dispersions (or b-values), and column densities for Fe II in the
$-$146 \kms{} system and \ion{Ti}{2} in the $-$513 \kms{}. The stellar
spectrum of \ecar\ is very complex due to multiple P-Cygni-shaped
profiles of iron and iron-peak singly-ionized elements. Equivalent
widths were obtained by  a best estimate of the stellar 'continuum'.
Estimated errors were obtained assuming reasonable limits on the
range of possible 'continua'. Thus the error range is much larger
than what would be expected for photon statistics. For the $-$146
\kms{} component, \ion{Fe}{2} was chosen for its large number of
spectral lines with equivalent widths lying on the linear portion
of the curve-of-growth (Table 1 and Figure 3). The $-$513 \kms{}
component is often blended with other components having velocities
between $-$385 and $-$585 \kms{}, making it more difficult to find
unblended lines that apply to curve-of-growth analysis (see Figure
2). However, many \ion{Ti}{2}, \ion{V}{2} and \ion{Fe}{1} lines
are unblended and have proven useful for analysis of the $-$513
\kms{} component. For this discussion we present curves-of-growth
for \ion{Ti}{2} for the $-$513 \kms{} component (Table 2 and
Figure 4).

\indent At this point, we must limit our analysis to those species
that have adequate gf-values and that have lines which can provide
reliable measurements to perform curve-of-growth analysis. We are
collaborating with several groups to obtain both experimental and
theoretical gf-values for the lines observed for other species
observed in the circumstellar spectrum of \ecar.
Consequently, a comprehensive, in depth analysis of these features
and the species they represent must wait until those data are
available.

\indent Variability of the physical conditions is of interest and
concern for this ejecta. As noted above, the spectra were obtained
in two visits separated by three months. Differences in the
equivalent widths were noticed for \ion{Fe}{2} lines ($-$146 \kms{} component) in the
overlap region in the 2900 \AA\ range. For that reason, we chose
to exclude measures of \ion{Fe}{2} lines from the January spectrum
and to use only measured lines up to 2940 \AA\ as recorded in the
October spectrum. We did not detect variations in equivalent
widths for the $-$513 \kms{} absorptions between October 2001 and
January 2002.
Hence we used equivalent width measures for \ion{Ti}{2} lines from
both the October 2001 and January 2002 spectra. Variability of the  $-$513 \kms{} component with
time is  occurring, but on a much longer timescale than for
the $-$146 \kms{} component. The $-$513 \kms{} component is found to be cooler,
 it is  likely further from the star and more shielded by intervening ejecta. This will be discussed in a later
paper describing changes from the broad maximum beginning in
October 2001 across the 2003.5 Minimum to the early stages of
recovery in March 2004.

\indent We measured equivalent widths for well-isolated
\ion{Fe}{2}  and \ion{Ti}{2} spectral lines and used them to
construct standard curves-of-growth. Column densities were
obtained for each energy level using transitions with detectable
absorption. The curves-of-growth for the three lowest terms of Fe II (Figure 3) are well represented by a
curve-of-growth with a b-value of 5.5 \kms{}, while the \ion{Ti}{2}
curves-of-growth for the two lowest terms (Figure 4) yield
b = 2.1 \kms{}. The good agreement further attests to the quality
of the equivalent width measurements and of the adopted gf-values.
This is most dramatically seen in the curve-of-growth for the
a$^4$D term in Fe II and in the a$^4$F term of
\ion{Ti}{2}. The very narrow range in the measured velocity centroids
($-$512.1 $\pm$ 1.0 and $-$145.8 $\pm$ 1.0 \kms), with no
systematic variations for lines with larger gf-values and
equivalent widths, imply negligible contribution from nearby
velocities.

\indent Table 3 presents the derived $-$146 \kms{} \ion{Fe}{2}
column densities, plus an estimate for  intermediate (14,000
to 21,000 cm$^{-1}$) energy levels not accessible in this spectral
region. The $-$513 \kms{} \ion{Ti}{2} has measurable populations
for the two lowest configurations (Table 4). The deduced falloff
in level populations with increasing excitation strongly suggest
that contributions from higher energy levels to the total column
densities  are negligible. In this spectral range, we find little
evidence for \ion{Ti}{1} or \ion{Ti}{3} for the $-$513 \kms{}
component nor for \ion{Fe}{1} or \ion{Fe}{3} for the $-$146 \kms{}
component. Thus, it is reasonable to assume that there are only
trace amounts of Ti and Fe in other ionization states. This allows us
to  convert derived column densities directly into total
elemental column densities.

\indent Because the lowest-lying levels are connected by only
parity-forbidden transitions, their populations should be well
within the conditions approaching LTE for anticipated densities
(see Verner et al. 1999, 2002). The critical densities for the departure coefficients departing from unity is N$_{\rm crit}$ =
A$_{ji}$/C$_{ji}$ $\approx$ 10$^6 -$ 10$^8$ cm$^{-3}$ ( Viotti 1976; see also Figure 3 in Verner et al. 1999). Thus, we can use the relative populations of the lowest-lying levels and
the Boltzmann Equation to estimate the characteristic temperature
in the line-forming regions of \ion{Fe}{2} and \ion{Ti}{2} for the
$-$146 \kms{} and the $-$513 \kms{} components, respectively. For
the higher energy states (E$_k$), even though they are coupled by
forbidden transitions, we would expect that their LTE departure
coefficients  to differ from unity. The actual departure coefficients depend strongly on the radiation field and cross-sections corresponding to permitted transitions connecting the lower metastable levels with higher non-metastable levels. The strong upward radiative rates and subsequent decays can push the departure coefficients for the higher energy metastable levels above unity. In actuality, the lower level populations ($<$26,000 \cm)
obey the Boltzman Equation such that the relative populations are
sensitive only to the electron temperature, T$_\mathrm{e}$.  Once T$_\mathrm{e}$ is
fixed by the relative populations of the lower levels,  the
departure coefficients of the higher energy levels become a
function of density and radiation field  (Verner et al. 1999).

\indent We obtained  estimates of the electron temperatures,
T$_\mathrm{e}$, by fitting the Boltzmann equation to the populations of the
lowest energy levels of \ion{Fe} {2} and \ion{Ti}{2}. Figures 5 and 6
demonstrate our fitting procedure using the Boltzman Equation,
$N_{l}/N_{0} = (g_{l}/g_{0})e^{(\chi_{l}-\chi_{0})/kT_e}$ to
obtain the  electron temperature in each component. In deriving
these temperatures, we have only used the deduced relative level
populations (i.e. column density for each energy level) for the
lowest energy levels to determine temperature. Here, $\chi_{0}$
denotes the mean energy for levels in the ground terms (See Tables 3 and 4). For the \ion{Fe}{2} in the $-$146 \kms\  component, our models as presented below indicate that populations of energy levels are in LTE up through  26,000 cm$^{-1}$.  As one increases in energy beyond 26,000 cm$^{-1}$, the models show an increase  above unity for the departure coefficients. At energies approaching $\approx$40,000 \cm,  the departure coefficients are less than five. Above energies of $\approx$40,000 \cm,  the  metastable levels disappear and the spontaneous decay rates rapidly depopulate the levels. The departure coefficients then approach zero. The net result of our modeling is that it appears that our relative populations of \ion{Fe}{2} is a function of temperature only. The density constraint that we can impose is that the total hydrogen particle density must be log(n$_\mathrm{H}$) $\ge$ 7 to 8.

\indent For the $-$146 \kms{} component,
T$_\mathrm{e}$ = 6400 K, (v$_{\rm therm}$ = 1.5 \kms, compared to a b-value
of 5.5 \kms{}), far below the implied temperature assuming the
velocity broadening - deduced from the b-value - is purely thermal
in origin. This indicates significant turbulence, or possibly
subcomponents, within each velocity system. In the case of the
$-$513 \kms{} component, the level populations of \ion{Ti}{2}, which we have assumed the level populations are represented by the Boltzman distribution,  are
consistent with T$_\mathrm{e}$ = 760 K (v$_{\rm therm}$ =0.5 \kms,
compared to a b-value of 2.1 \kms{}).

\subsection{Modeling the Velocity Components}
\indent We have applied photoionization simulations, incorporating
statistical  equilibrium calculations for a 371-level model for
Fe$^+$ to produce a grid of models representing the absorbing gas
of the the $-$146 and the $-$513 \kms{} components. Similar
simulations have been used successfully to model the Fe II and [Fe
II] emission arising in the Weigelt Blobs B \&\ D (Verner et al.
2002) - ejecta of $\eta$ Car associated with the 1890's 
event (Smith et al 2004). The models in our grid are calculated
based upon measured column densities, temperatures, and relative
level populations derived from the curve-of-growth analysis. From
our previous work (Verner et al. 2002), we  adopt a stellar luminosity of 10$^{40}$ergs$^{-1}$ with
T$_{\rm eff}$ = 15,000 K. We approximate the stellar flux
distribution by a Kurucz (1979, 1988) model atmosphere. We assume that
the ejecta have undergone CNO-processing, where carbon and oxygen
are at 0.01 solar abundance, with nitrogen at 10x solar abundance
(Verner, Bruhweiler, \& Gull
2004; also see Hillier et al. 2001).  Given our poor overall knowledge
about dust formation processes in the Eta Carinae environment,
uncertainties in gas-phase fraction, dust particle size and its
chemical compositions, the effects of dust are not included into
our current calculations. However, the geometry of the Little Homunculus indicates that a hot, wind-blown cavity is between the star and the Little Homunculus (Ishibashi et al. 2003, Nielsen et al. in prep). We thus expect little or no dust between the central source and the $-$146  \kms\ component. 

\indent  The temperature
of the low velocity  ($-$146 \kms{}) component is very close to
that previously derived for the Weigelt Blobs B and D (Verner et al. 2002).
Thus, it is reasonable to expect an ionization for this component to be  similar to the Weigelt
Blobs. Unfortunately, our spectral data do not
provide a means to estimate the  (Fe/H) abundance ratio.  On the
other hand our emission line study (Verner, Bruhweiler and Gull
2004) as well as the investigation by Hillier et al. (2001) of the
stellar spectrum show that (Fe/H) ranges from 0.2$\times$(Fe/H)$_\mathrm{solar}$  in the BD Blobs to
~2$\times$(Fe/H)$_\mathrm{solar}$, respectively. Therefore given the uncertainties, we have
adopted the (Fe/H) to be (Fe/H)$_\mathrm{solar}$   in our calculations.

\indent The most accurate measurements of the relative Fe II level populations (or column densities) are  for
the second (\term a4F) and third (\term a4D) configurations. We have constructed the
theoretical level population ratio and scaled the other Fe II
level populations derived from the curves-of-growth to this ratio.
The Fe II relative energy level population ratio (R=n(\term a4D)/n(\term a4F)) is plotted
in Figure 7 for a range of hydrogen densities from 10$^5$ to
10$^8$cm$^{-3}$ and
 a range of
stellar distance between 10$^{15.5}$ and 10$^{16.5}$cm. The best  models suggest the temperature range is from 5700 K to 7300 K, well bracketing the \ion{Fe}{2}-inferred temperature of  6400 K (Figure 5).  The
predicted \ion{Fe}{2} populations  for the a$^4$F and a$^4$D terms are found to be in LTE.  In Figure 8 the relative \ion{Fe}{2} population levels are compared to the predicted level populations for n$_\mathrm{H} = 10^7$ and $10^8$cm$^{-3}$ at distances 10$^{16}$  and 10$^{16.5}$ cm.   From Figure 7, the normalized (unity) ratio indicates the stellar distance for the $-$146 \kms{} to be
 10$^{16.3}$ cm, or  $\approx$1300 AU.

\indent  Independently, we repeated calculations for the $-$513
\kms{}  component. We  measured only column densities for \ion{Ti}{2}
from the two lowest configurations as indicated above.
Consequently, we cannot derive total hydrogen densities for this
absorbing region. However, given the low deduced T$_\mathrm{e}$ and low
ionization of this component, one must conclude that if it were at
a distance comparable or less than the $-$146 \kms{} component, it
would necessarily have a higher density. Otherwise, the ionization
and temperature would be much higher than the T$_\mathrm{e}$ = 760 K found.
From a purely qualitative argument, we conclude the $-$513 \kms{}
component, unless it has exceptionally high densities, is at a
distance much farther than 1300 AU. 

\indent For convenience, we adopt a n$_\mathrm{H}\ge $10$^7$cm$^{-3}$, similar
to that found for the –146 \kms\ component.  Again, we assume the
same stellar luminosity, but use the temperature derived from the
\ion{Ti}{2} energy populations, T$_\mathrm{e}$=760 K. Our
photoionization modeling then places the  $-$513 \kms{} component
eight times more distant, or r$\approx$10,000 AU from the stellar
source.

\indent Based upon our modeling, we present the deduced physical
conditions for the $-$146 \kms {} and the $-$513 \kms{} components
in Table 5. Of course, these results are somewhat dependent upon
the gaseous elemental depletions for Fe ($-$146 \kms
{}) and the Ti ($-$513 \kms{}). We have also
ignored possible shielding in these calculations. For example, if
other components, lying between the central stellar source and the
$-$146 \kms{} and $-$513 \kms, produce significant attenuation of the
\ecar\ radiation field, then this would reduce the implied
distance for that component. 

\section{Interpretation}
The geometry of the Homunculus, based upon STIS CCD long aperture
spectra of nebular emission and absorption lines originating from
the interior surface of the expanding debris, has been determined
to be a bipolar lobed structure at an inclination angle of
$\sim$41$^o$ degrees with respect to the plane of the sky
(Davidson et al. 2001). Spatial variations along the slit are
resolved at the $ 0\,\farcs1$ \ level, a projected scale of about
230 AU at the distance of \ecar. The line-of-sight, centered on
\ecar{}, passes almost tangentially through the foreground lobe of
the Homunculus, and likely intersects clumps of corresponding
size. The Homunculus is  a relatively thin shell. However, the
analysis of Davidson et al (2001) used the emission of
[\ion{Fe}{2}] and [\ion{Ni}{2}] lines to describe what appeared to
be a relatively diffuse, but thin interior surface of the
Homunculus. N. Smith (private communication) comments that the
infrared spectroscopy of the Homunculus is consistent with a shell
about one-tenth the thickness of the distance from \ecar, and with
a relatively diffuse interior skin and very thin outer skin.

\indent The $-$513 \kms{} component is likely part of the
Homunculus in the line-of-sight. Based upon its relatively small
b-value, it may be associated with the thin outer skin. Likewise, the
detected components with velocities between $-$385 and $-$585
\kms{} are evidence for clumps within the relatively thick wall of
the Homunculus. Interior to the Homunculus is a large cavity,
filled with a hot, very low density gas, too faint to be detected by
current observations. Still interior is the Little Homunculus
(Ishibashi et al. 2003) that is a thin, ionized shell with
measured velocities in the vicinity of $-$146 \kms{}. Expansion
velocities and proper motions of both the Homunculus and the
Little Homunculus are consistent with ejection during the major
outburst in the 1840's and the lesser outburst in the 1890's. The
central region, immediately surrounding \ecar{}, is thought to be
a wind-blown cavity, with an extended skirt whose plane is
perpendicular to the axis of the two lobes. In this picture, the
two major velocity components seen in the STIS NUV spectra arise in the
gas ejected during each of the two outbursts. The other velocity
components, intermediate in velocity space, likely are components
of the outer Homunculus, distributed in that one-tenth thickness,
but projected along a long line of sight due to geometry. We
note in passing that these multiple components are photoionized 
(or excited) by \ecar, and that  during the minimum, very
significant changes occur.  Data have been obtained during the
2003.5 minimum with the STIS, analysis on the temporal variability
is ongoing.

\indent The presence of dust and its effects on the excitation of these ejecta components appears to be surprisingly little. However, given the hot, low-density cavities bounded by the Little Homunculus and by the interior wall of the Homunculus, little dust intervenes between the star and the Little Homunculus, and in turn between the Little Homunculus and the outer Homunculus. The bulk of the dust is likely contained in the clumps of ejecta in line of sight. Hence we might expect to see some dust contribution in the thin wall of the Little Homunculus ($-$146 \kms{} component) and considerably more dust contributions in the thicker wall of the Homunculus. Indeed dust is present in the Homunculus and is the scatterer for the stellar radiation seen reflected of the bipolar lobes. Smith et al (2003) demonstrated that the Balmer alpha (H$\alpha$) P-Cygni line profile changed across the 5.54-year cycle consistent with the terminal wind velocity increasing to about 1200 \kms{} during the maximum and dropping to about 600 \kms{} during the minimum. These changes in wind velocity are interpreted to originate from the high latitudes of the central source and that the dust in the relatively flat polar regions of the Homunculus has an optical depth of order unity, therefore causing a single scattering event between the star and the observer. Given the very lumpy appearance of the flat polar region and the relatively linear striations of the walls of the Homunculus, dust appears to be clumpy in distribution in the outer structure of the Homunculus. The simplest interpretation is that most of the dust in line of sight resides in the outer wall of the Homunculus and that some may lie in the interior thinner wall of the Little Homunculus. Certainly the presence of \ion{Fe}{2}  absorption lines in the Little Homunculus and \ion{Fe}{1} absorption lines in the Homunculus indicates that much of the FUV radiation is stopped by the Little Homunculus. Some FUV radiation down to the Lyman limit does get through the wall as the STIS FUV and FUSE spectra show complex, but relatively continuous UV  radiation to well  below 1000 \AA\ originating from a nearly point-like source. However no locally emitted Balmer line emission is detected from the Homunculus, but near-red [\ion{Fe}{2}] and [\ion{Ni}{2}] emission lines are present and were used by Davidson et al. (2001) to trace the geometrical structure of the Homunculus. The emission line velocities of the Little Homunculus and the Homunculus in the line of sight are in good agreement with the absorption velocities we see. Thus, the simplest interpretation of these absorption components are that they are associated with the expanding ejecta, not clumps within the much hotter wind within a few hundred AU of the star.

\indent Many of the elements identified in these velocity
components are refractory, with condensation temperatures below
1300 K. In a later paper we plan to address relative abundances,
but initial inspection indicates that elemental depletion is low
and not following a Spitzer-Routly law (Routly \& Spitzer 1952).
In particular the presence of \ion{V}{2} and \ion{Ti}{2} at $-$513
\kms{} is remarkable.  We speculate that the combination of low C
and O  abundances, the expanding ejecta and/or the modulated wind
may inhibit normal grain formation in the \ecar\ system.

\section{Conclusions}
The UV STIS spectrum of \ecar{} reveals complex multiple, narrow
absorption features for many neutral and singly-ionized species,
spanning a velocity range from $-$146 \kms{} to $-$585 \kms{}. Two
narrow velocity components ($-$146 and $-$513 \kms{}) appear to
originate from shells or clumps in the foreground lobe of the
Homunculus. These clumps are heated by the very bright UV flux of
\ecar. The presence of ionic transitions up to energies of 40,000
cm$^{-1}$ is, in part, due to the high density in ejecta giving
rise to the observed very strong circumstellar absorptions. The
photoionization modeling for the $-$146 and the $-$513 km~s$^{-1}$
components suggest that the corresponding absorptions arise at large
distances from the central stellar source. We conclude that the
absorption is formed in dense gas located at different
positions within the bipolar structure of the Homunculus. The
$-$146 \kms{} component is formed at a distance of $\sim$1300 AU,
which leads to a spectrum consisting mainly of lines from
singly-ionized iron-group ions. The $-$146 \kms{} component
is likely formed in the wall of the Little Homunculus (Ishibashi et al.
2003). Although the conclusion is less firm,  the cooler $-$513
\kms{} component, seen in the lines of neutral and singly ionized
elements, appears to be formed at a much larger distance from the
star ($\sim$10,000AU). This component is likely formed in the wall
of the Homunculus. Further analysis of the STIS dataset for \ecar,
including changes that occur during the 2003.5 minimum will be
presented in subsequent papers.

\acknowledgments{This paper is based upon observations made with
the NASA/ESA Hubble Space telescope, obtained at the Space
Telescope Science Institute, which is operated by the Association
of Universities for Research in Astronomy, Inc., under NASA
contract NAS 5-26555.} Funding was provided through the STIS
Guaranteed  Time Observations (GTO). The research of EV has been supported through STIS GTO funding and through  NSF - 0206150 to CUA.

\begin{deluxetable}{lrrcrrrc}
  \tabletypesize{\scriptsize}
\tablecaption{\ion{Fe}{2} lines used in the $-$146 \kms{}
component analysis \label{fe2tab}} \tablewidth{0pt} \tablehead{
\colhead{Lower Level} &
\colhead{$\lambda_\mathrm{lab}$\tablenotemark{a}} &
\colhead{Vel.\tablenotemark{b}} &
\colhead{J$_\mathrm{low}$\tablenotemark{c}} &
\colhead{E$_\mathrm{low}$\tablenotemark{d}} &
\colhead{W$_\lambda$\tablenotemark{e}} &
\colhead{log\,($gf$)\tablenotemark{f}} &
\colhead{Ref.} \\
\colhead{}& \colhead{(\AA)}& \colhead{(km\,s$^{-1}$)}& \colhead{}&
\colhead{(cm$^\mathrm{-1}$)}& \colhead{(\AA)}& \colhead{}&
\colhead{} } \startdata
($^{5}$D)4s a$^{6}$D& & & && & \\
& 2586.650  &$-$143.8 &  4.5 &   0.000 &0.224 & $-$0.19 &FMW\\
& 2600.173  &$-$145.9&  4.5 &   0.000 &0.264 &    0.35 &FMW\\
& 2599.147  &$-$145.2&  3.5 & 384.790 &0.165 & $-$0.10 &FMW\\
& 2399.973  &$-$144.8&  2.5 & 667.683 &0.184  & $-$0.15 &FMW\\
& 2407.394  &$-$144.9& 1.5  & 862.613 &0.176 & $-$0.25 &FMW\\
& 2614.605  &$-$142.9& 1.5  & 862.613 &0.180 & $-$0.39 &FMW\\
& 2621.191  &$-$146.0& 1.5  & 862.613 &0.050 & $-$1.83 &FMW\\
& 2414.045  &$-$145.2& 0.5  & 977.053 &0.165 & $-$0.43 &FMW\\
& 2622.452  &$-$146.1& 0.5  & 977.053 &0.119 & $-$1.00 &FMW\\
d$^{7}$ a$^{4}$F    && & & && & \\
& 2451.842  &$-$146.4& 4.5 & 1872.567 &0.031& $-$2.07 & K88\\
& 2484.951  &$-$146.5& 4.5 & 1872.567 &0.047& $-$1.98 & K88\\
& 2392.206  &$-$146.7& 3.5 & 2430.097 &0.056& $-$1.64 & FMW\\
& 2505.969  &$-$147.0& 3.5 & 2430.097 &0.022& $-$2.22 & K88\\
& 2512.125  &$-$146.6& 3.5 & 2430.097 &0.018& $-$2.53 & K88\\
($^{5}$D)4s a$^{4}$D& && & && & \\
&2563.304 &$-$144.7& 3.5 &  7955.299 &0.156& $-$0.05 & FMW\\
&2693.633 &$-$146.5& 3.5 &  7955.299 &0.026& $-$2.09 & FMW \\
&2715.218 &$-$146.2& 3.5 &  7955.299 &0.256& $-$0.44 & FMW\\
&2740.358 &$-$145.8& 3.5 &  7955.299 &0.203& 0.24 & FMW \\
&2756.551 &$-$144.8& 3.5 &  7955.299 &0.224&  0.38   & FMW\\
&2881.601 &$-$145.5& 3.5 &  7955.299 &0.041& $-$1.66 & FMW\\
&2927.441 &$-$146.9& 3.5 &  7955.299 &0.085& $-$1.23 & FMW\\
&2564.245 &$-$146.0& 2.5 &  8391.938 &0.151& $-$0.29 & FMW\\
&2869.716 &$-$147.0& 2.5 &  8391.938 &0.012& $-$2.29 & FMW\\
&2940.367 &$-$146.7& 2.5 &  8391.938 &0.008& $-$2.75 & FMW\\
&2567.683 &$-$145.3& 1.5 & 8680.454 &0.098& $-$0.65 & FMW\\
&2731.542 &$-$145.7& 1.5 & 8680.454 &0.082& $-$0.95 & FMW\\
&2769.753 &$-$146.4& 1.5 & 8680.454 &0.047& $-$1.51 & FMW\\
&2762.629 &$-$147.1& 0.5 & 8846.768 &0.068& $-$1.28 & FMW\\
\enddata
\tablenotetext{a}{\ Laboratory wavelengths} \tablenotetext{b}{\
Measured velocity of absorption line} \tablenotetext{c}{\ J-values
for the lower energy state} \tablenotetext{d} {\ Energy of the
lower state} \tablenotetext{e} {\ Measured equivalent width}
\tablenotetext{f} { \  log(gf) values from references}

\tablerefs{ (FMW) $-$ Fuhr J.~R., Martin G.~A, and Wiese W.~L.
1988, J.~Phys. Chem. Ref. Data 17, Suppl. 4, (K88) $-$ Kurucz
R.~L. 1988, In: McNally M. (ed.) Trans. IAU, XXB, Kluver,
Dordrecht, p. 168 }
\end{deluxetable}

\begin{deluxetable}{lrrcrrr}
 \tabletypesize{\scriptsize}
\tablecaption{\ion{Ti}{2} Lines used in $-$513 \kms{} Component
Analysis \label{ti2tab}} \tablewidth{0pt} \tablehead{
\colhead{Lower Level} &
\colhead{$\lambda_\mathrm{lab}$\tablenotemark{a}} &
\colhead{Vel.\tablenotemark{b}} &
\colhead{J$_\mathrm{low}$\tablenotemark{c}} &
\colhead{E$_\mathrm{low}$\tablenotemark{d}} &
\colhead{W$_\lambda$\tablenotemark{e}} &
\colhead{log\,($gf$)\tablenotemark{f}} \\
\colhead{}& \colhead{(\AA)}& \colhead{(km\,s$^{-1}$)}& \colhead{}&
\colhead{(cm$^\mathrm{-1}$)}& \colhead{(\AA)}& \colhead{} }
 \startdata
($^{3}$F)4s a$^{4}$F& & & &&  \\
&3058.282&$-$513.1 &1.5  &  0.00    &0.016 & $-$1.78  \\
&3067.238&$-$513.5 & 1.5  &  0.00    &0.053& $-$0.71  \\
&3073.863&$-$512.5 &    1.5  &  0.00    &0.070 & $-$0.32  \\
&3122.515 &$-$514.4 &   1.5 &   0.00    &0.006 & $-$2.36  \\
&3148.948&$-$513.0 &    1.5  &  0.00    &0.043& $-$1.22  \\
&3060.634 &$-$513.9 &   2.5 &   94.10   &0.012& $-$1.57  \\
&3067.109&$-$513.3 &    2.5 &   94.10   &0.052& $-$0.59  \\
&3076.117&$-$512.5 &    2.5 &   94.10   &0.080& $-$0.12 \\
&3131.706&$-$513.0 &    2.5 &   94.10   &0.026& $-$1.19 \\
&3158.307&$-$513.7 &    2.5 &   94.10   &0.008& $-$2.17 \\
&3073.000&$-$512.9 &    3.5  &  225.73  &0.046& $-$0.62 \\
&3079.539&$-$512.6 &    3.5  &  225.73  &0.081& 0.06 \\
&3144.665&$-$512.4 &    3.5  &  225.73  &0.019& $-$1.30 \\
&3088.922&$-$512.1 &    4.5 &   393.44  &0.085&    0.23 \\
d$^{3}$ b$^{4}$F  && & & & & \\
&2536.632&$-$512.1 &    1.5 &   908.02  &0.009 & $-$1.03 \\
&3155.106&$-$512.2 &    1.5 &   908.02  &0.016 & $-$1.15  \\
&3162.117&$-$512.1 &    1.5 &   908.02  &0.023 & $-$0.69\\
&2535.381&$-$512.9 &    2.5 &   983.89  &0.012 & $-$0.94\\
&3153.163&$-$510.7 &    2.5 &   983.89  &0.012 & $-$1.06\\
&3162.684&$-$512.0 &    2.5 &   983.89  &0.032 & $-$0.55\\
&2532.013&$-$512.7 &    3.5 &   1087.32 &0.012 & $-$0.67\\
&3156.582&$-$510.9 &    3.5 &   1087.32 &0.008 & $-$1.17\\
&3163.481&$-$512.1 &    3.5 &   1087.32 &0.033 & $-$0.38\\
&2526.362&$-$512.8 &    4.5 &   1215.84 &0.014 & $-$0.51\\
\enddata
\tablenotetext{a}{\ Laboratory wavelengths S. Johansson (private
communication)} \tablenotetext{b}{\ Measured velocity of
absorption line} \tablenotetext{c}{\ J-values for the lower energy
state} \tablenotetext{d} {\ Energy of the lower state}
\tablenotetext{e} {\ Measured equivalent width} \tablenotetext{f}
{\ log(gf) values from Pickering, Thorne \& Perez (2001,2002) }

\end{deluxetable}

\begin{deluxetable}{lccc}
  \tabletypesize{\scriptsize}
  \tablecaption{Column densities from the $-$146 \kms{} 
  \ion{Fe}{2} curve-of-growth analysis. }
\tablewidth{0pt} \tablehead {
 \colhead{ } &
 \colhead{Conf.} &
 \colhead{E$_{\mathrm{mean}}$(\cm)}\tablenotemark{1} &
  \colhead{n$_{i}$(10$^\mathrm{13}$ cm$^{-2}$)}  }
  
\startdata
&\conf 4s\, \term a6D & 416 & 169.9  \\
&\conf 3d7 \term a4F & 2417 & 200.0  \\
&\conf{4}{s}{} \term a4D & 8321 &  54.56 \\
&\term a4P \tablenotemark{2} & 18007 &  59 \tablenotemark{3}  \\
&\conf {4}{s}{} \term b4P & 21421 & 0.83\\
&\conf{4}{s}{} \term a4H & 21459  &  3.85  \\
&\conf{4}{s}{} \term b4F & 22808 & 2.34  \\
&\conf{4}{s}{} \term a4G & 25761 & 0.95  \\
&\conf{4}{s}{} \term b2P & 26169 & 0.51 \\
&\conf{4}{s}{} \term b2H & 26253 & 0.99 \\
&\conf{4}{s}{} \term a2F & 27445 & 0.86 \\
&\conf{4}{s}{} \term b2G & 30556 &0.42 \\
&\conf{4}{s}{} \term b4D & 31421 & 0.68 \\
&\conf{4}{s}{} \term a2I  & 32892 & --  \\
&\conf{4}{s}{} \term c2G & 33482 & 0.18 \\
&\conf{4}{s}{} \term c2D & 38184 &  0.28 \\ \noalign{\smallskip}
N =$ \sum_{i} { \mathrm{n} _i}$\tablenotemark{4} &&& 495 \\
\enddata
\tablenotetext{1}{E$_{\mathrm{mean}}$=$ \sum_{i} {( \mathrm{E} _i}\mathrm{g}_{i})/\sum_{i}g_i$}
\tablenotetext{2}{\ Missing energy states intermediate in energy between \conf{4}{s}{}\term
a4D and \conf 3d7 \term b4P} \tablenotetext{3}{Estimated column
density based on interpolation}\tablenotetext{4}{For completeness, we also note that column densities for energy states \term a6S, \term b2F, \term b2D, and \term a2S were also not measured but  as at high energy levels would contribute a negligible amount to the total population.}
\end{deluxetable}

\begin{deluxetable}{lccc}
  \tabletypesize{\scriptsize}
  \tablecaption{ Column densities from the $-$513 \kms{} \ion{Ti}{2}
curve-of-growth analysis.} \tablewidth{0pt} \tablehead{ \colhead{
} & \colhead{Conf.} & \colhead {E$_{\mathrm{mean}}$(\cm)}\tablenotemark{1} & 
\colhead{n$_{i}$ (10$^\mathrm{13}$ cm$^{-2}$)}} \startdata
&\conf 4s\, \term a4F & 225 & 16 \\
&\conf 3d3 \term b4F & 1085 & 3 \\
\noalign{\smallskip}
N = $\sum_{i} {\mathrm{n}_i}$ &&& 19   \\
\enddata
\tablenotetext{1}{E$_{\mathrm{mean}}$=$ \sum_{i} {( \mathrm{E} _i}\mathrm{g}_{i})/\sum_{i}g_i$}
\end{deluxetable}

\begin{deluxetable}{lcc}
\tabletypesize{\scriptsize}
\tablecaption{Physical parameters for the $-$146 and $-$513 \kms
components \label{table1}} \tablewidth{0pt} \tablehead{ \colhead{
} & \colhead{$-$146 \kms{}} & \colhead{ $-$513 \kms{}}} \startdata
b-value& 5.5 \kms{} & 2.1 \kms{}\\
Derived T$_\mathrm{e}$: & 6400 K & 760 K \\
Ionization & Singly Ionized Species& Neutral \& \\
&& Singly Ionized Species\\
Hydrogen density, n$_\mathrm{e}$: & $\ge$10$^7$cm$^{-3}$ &
$\ge$10$^7$cm$^{-3}$ \\
Distance from star&$\sim$1300 AU & $\sim$10,000 AU \\

\enddata
\end{deluxetable}

\begin{figure}[ht]
\scalebox{0.8}  {\includegraphics{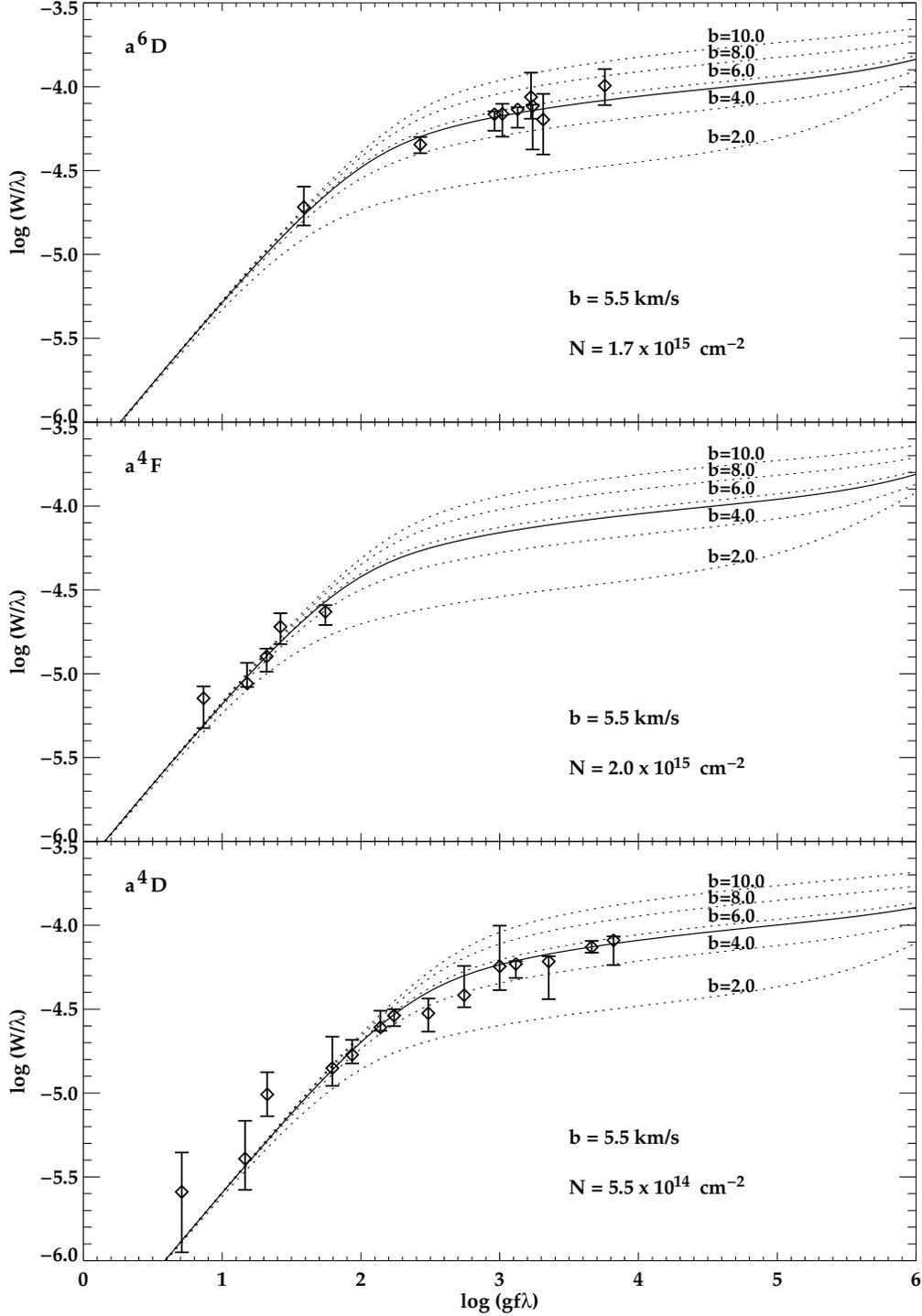}} \caption{ The $-$146 \kms{}
Component curves-of-growth plots for the lowest three terms of \ion{Fe} {2}. Measured lines were selected to minimize
blending with other velocity components, to be measurable against
the relatively complex stellar spectrum. The relatively large
uncertainty bars are due to ability to determine the continuum level of
\ecar, and not photon statistics. Note that all three
curves-of-growth are consistent with a b-value of 5.5 \kms{}.
 \label{fig3}}
\end{figure}

\begin{figure}[ht]
\scalebox{0.8}  {\includegraphics{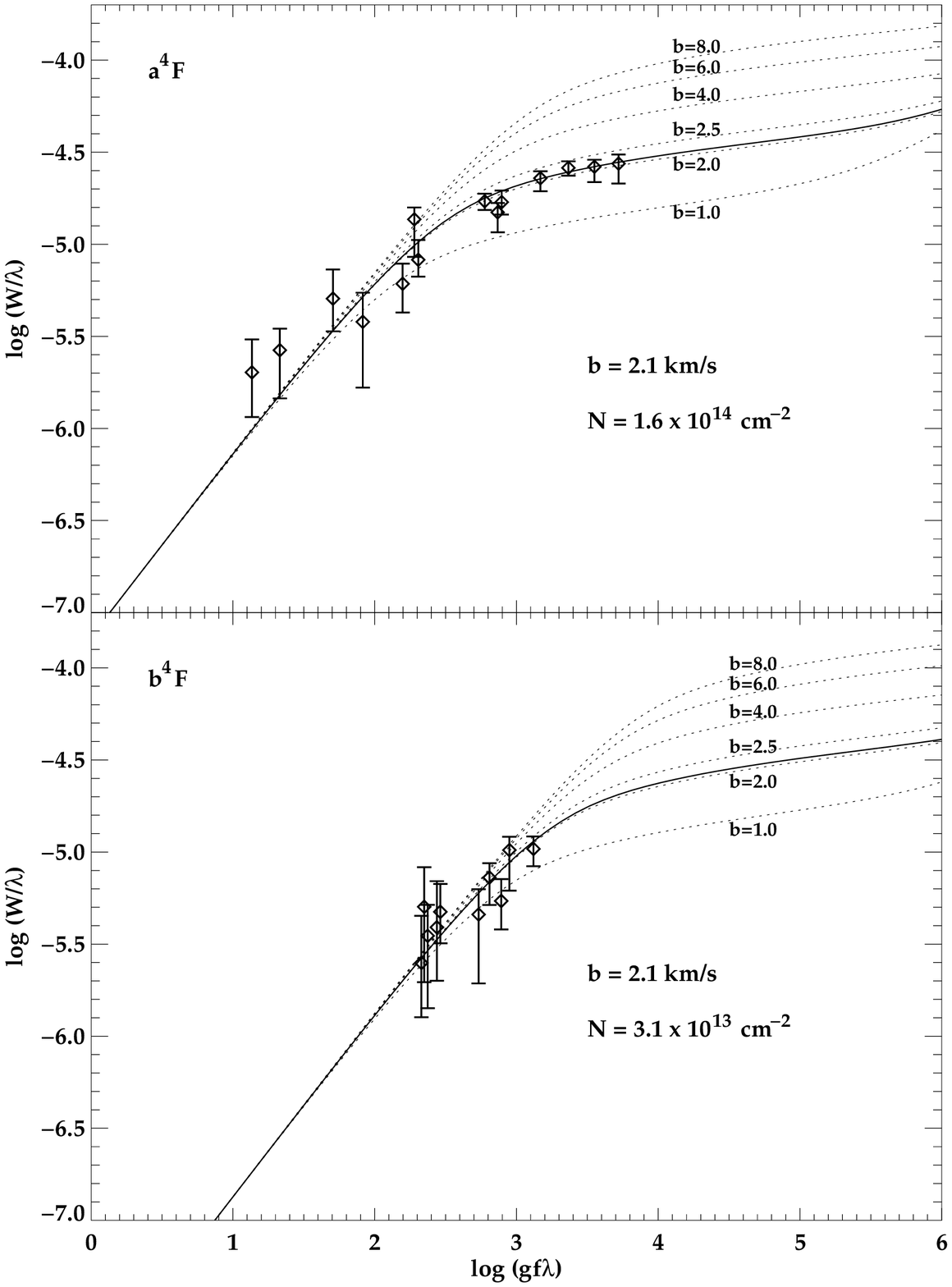}} \caption{The $-$513 \kms{}
Component curves-of-growth plots for the lowest two terms
of \ion{Ti}{2}. The $-$513 \kms{} component is much narrower in
width and is seen in lines originating from significantly lower
energy levels that seen either in the P-Cygni wind of \ecar\ or in
the multiple absorption components similar in excitation to the
$-$146 \kms{} component. No lines originating from higher energy
levels were found. Thus the populations of these two terms
are found to be the major contributors to the total amount of \ion{Ti}{2}.\label{fig4}}
\end{figure}

\begin{figure}[ht]
\scalebox{0.7}{\rotatebox{90}{\plotone{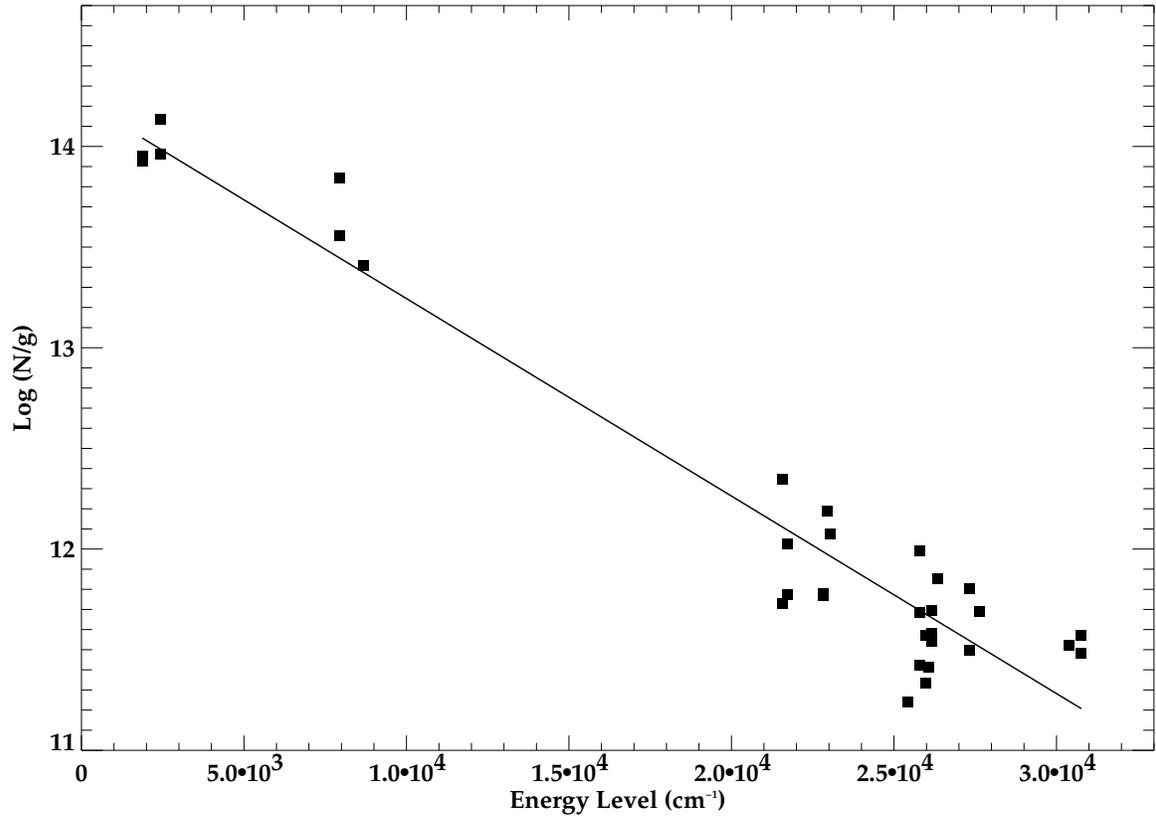}}} \caption{Fe II
population for each energy level of the $-$146 \kms{} component.
The linear slope corresponds to 6370 K. Note that the
population for each energy level is normalized by the statistical weight. \label{fig5}}
\end{figure}

\begin{figure}[ht]
\scalebox{0.7}{\rotatebox{90}{\plotone{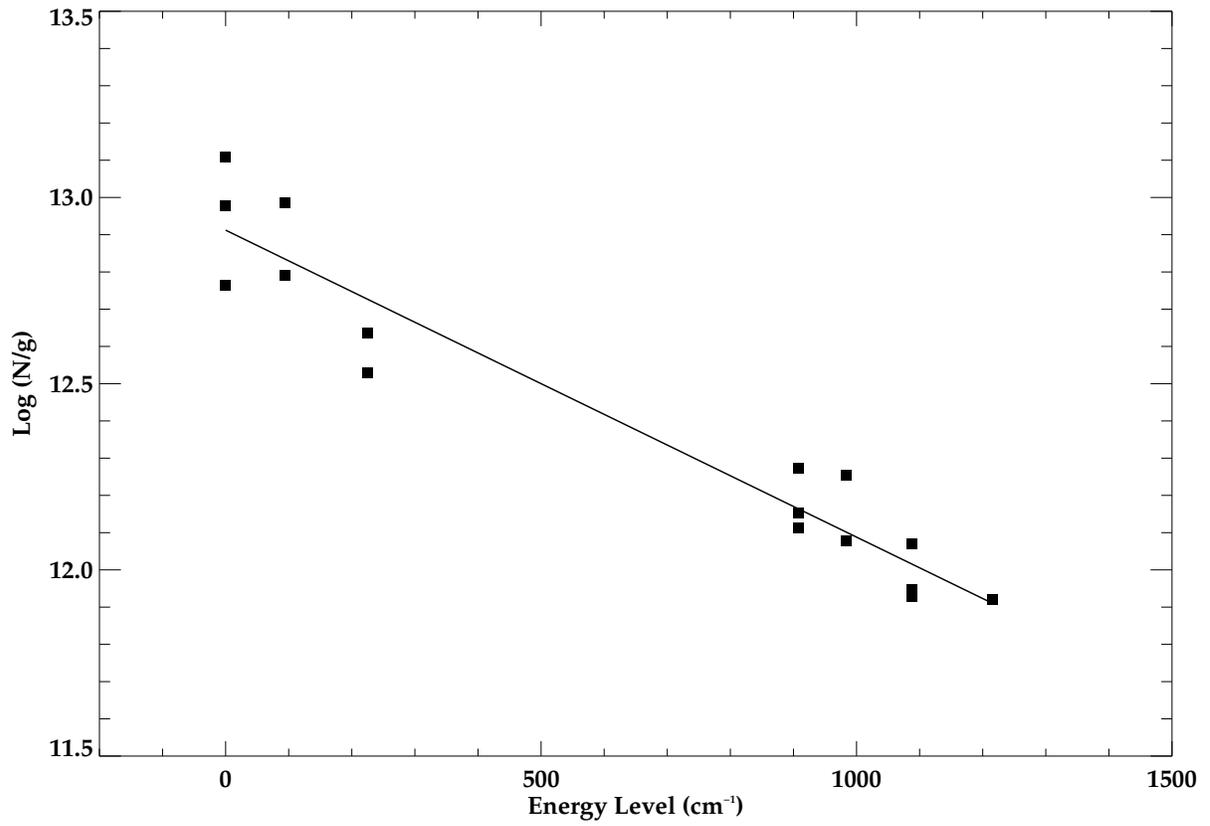}}} \caption{\ion{Ti}{2}
population for each energy level of the $-$513 \kms{} component.
The linear slope corresponds to 760 K.\label{fig6}}
\end{figure}

\begin{figure}[ht] 
\scalebox{1.} {\rotatebox{-90}{\plotone{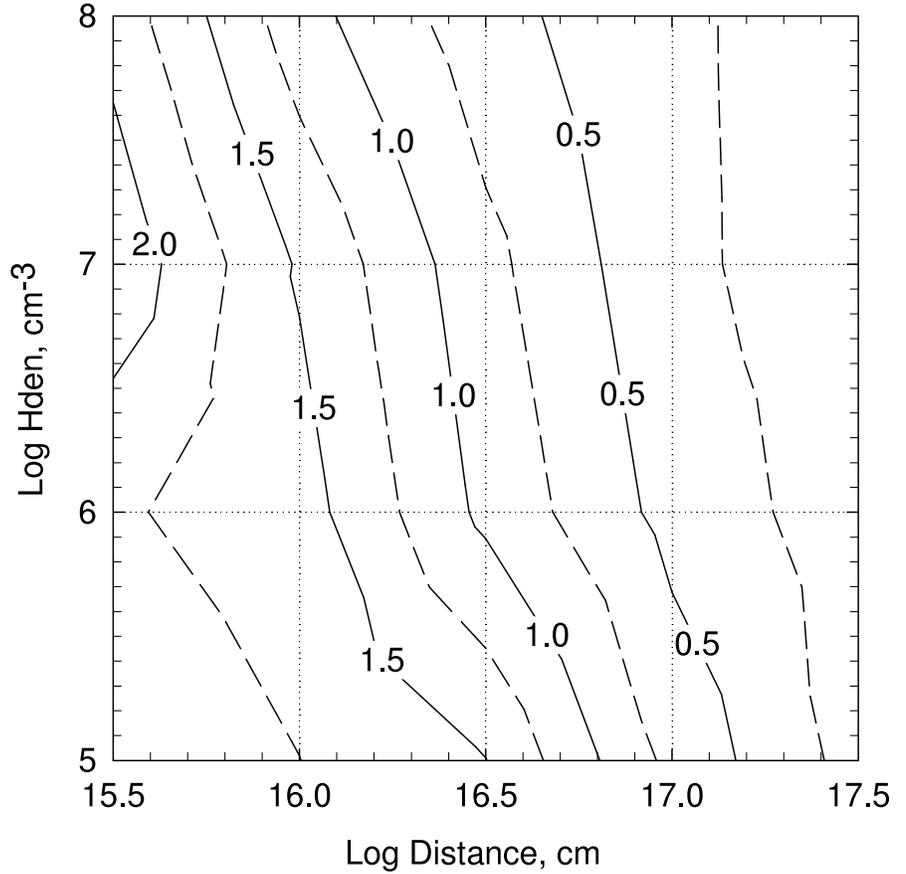}}} \caption{The $-$146 \kms{}
component: hydrogen density versus distance plot. The normalized ratio of the \ion{Fe}{2}
column densities for the a$^4$F and a$^4$D terms, R=(a$^4$D/a$^4$F)$_{\mathrm{theory}}$/(a$^4$D/a$^4$F)$_{\mathrm{observation}}$, was used to determine the optimal density/distance value. A value of unity for R indicates a match to the derived column density. At a
distance of 10$^{16.1}$ to 10$^{16.4}$ cm, the hydrogen column
density would be between 10$^7$ to 10$^8$ cm$^{-3}$.\label{fig7}}
\end{figure}

\begin{figure}[ht]
\  \scalebox{.8} {\rotatebox{-90}{\includegraphics{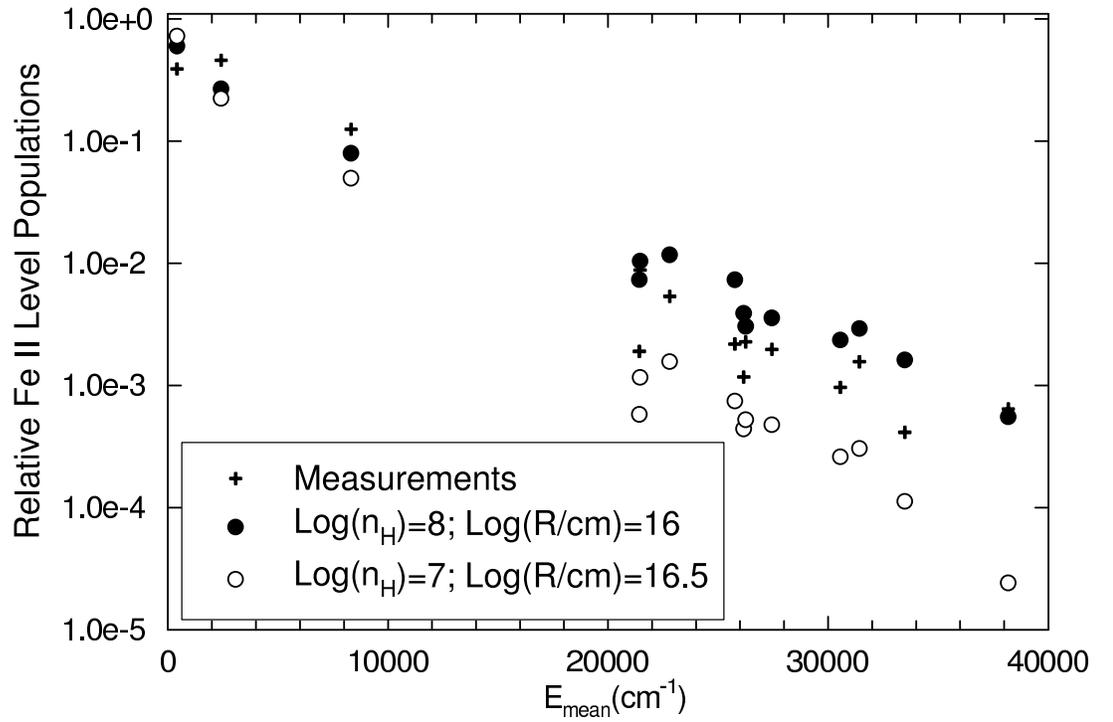}}} \caption{The
observationally-derived and predicted \ion{Fe}{2} relative 
level populations for the $-$146 \kms{} component. 
The best fit lies  between the two density values, possibly closer
to n$_\mathrm{H}$=10$^8$ cm$^{-3}$. The  theoretical and observed relative level populations roughly approximate a straight line for the three cases shown.  The  slope reflects the electron  temperature, T$_e$, for each dataset.  Most of the scatter is due to differing statistical weights for each level. \label{fig8}}
\end{figure}

\end{document}